\documentclass[12pt]{article}
\usepackage[english]{babel}
\usepackage{graphicx}               
\usepackage{amsmath,amssymb,algorithm,algorithmic,bbm,multirow,mathtools,booktabs,amsthm}

\usepackage[utf8]{inputenc}
\usepackage{comment}

\usepackage{natbib}
\usepackage{float}

\allowdisplaybreaks
\newcommand{\E}{\mathbb{E}}

\newtheorem{prop}{Proposition}

\usepackage[colorlinks,citecolor=blue,urlcolor=blue]{hyperref}
\newcommand{\blind}{1}

\begin{document}

\def\spacingset#1{\renewcommand{\baselinestretch}%
{#1}\small\normalsize} \spacingset{1}


\if1\blind
{
  \title{\bf Bayesian Factor Analysis for Inference on Interactions}
  \author{Federico Ferrari \\
    Department of Statistical Science, Duke University\\
    and \\
    David B. Dunson \\
    Department of Statistical Science, Duke University}
  \maketitle
} \fi

\if0\blind
{
  \bigskip
  \bigskip
  \bigskip
  \begin{center}
    {\LARGE\bf Bayesian Factor Analysis for Inference on Interactions}
\end{center}
  \medskip
} \fi

\bigskip
\begin{abstract}
This article is motivated by the problem of inference on interactions among chemical exposures impacting human health outcomes. Chemicals often co-occur in the environment or in synthetic mixtures and as a result exposure levels can be highly correlated. We propose a latent factor joint model, which includes shared factors in both the predictor and response components while assuming conditional independence. By including a quadratic regression in the latent variables in the response component, we induce flexible dimension reduction in characterizing main effects and interactions. We propose a Bayesian approach to inference under this Factor analysis for INteractions (FIN) framework. Through appropriate modifications of the factor modeling structure, FIN can accommodate higher order interactions. We evaluate the performance using a simulation study and data from the National Health and Nutrition Examination Survey (NHANES). Code is available on GitHub.
\end{abstract}
\noindent%

\newpage
\spacingset{1.5} 
\section{Introduction}
\label{sec:intro}

There is broad interest in incorporating interactions in linear regression. Extensions of linear regression to accommodate pairwise interactions are commonly referred to as quadratic regression. In moderate to high-dimensional settings, it becomes very challenging to implement quadratic regression since the number of parameters to be estimated is $2 p + {p \choose 2}$. Hence, classical methods such as least squares cannot be used and even common penalization and Bayesian methods can encounter computational hurdles. Reliable inferences on main effects and interactions is even more challenging when certain predictors are moderately to highly correlated. 

A lot of effort has been focused on estimating pairwise interactions in moderate high-dimensional and ultra high-dimensional problems. We refer to the former when the number of covariates is between $20$ and $100$ and to the latter when $p>100$. When $p = 100$, the number of parameters to be estimated is greater than $5000$. When $p \in [20,100]$, one-stage regularization methods like \cite{Bien2013} and \cite{Haris2018} can be successful. Some of these methods require a so-called heredity assumption \citep{Chipman1996} to reduce dimensionality. Strong heredity means that the interaction between two variables is included in the model only if both main effects are. For weak heredity it suffices to have one main effect in the model. Heredity reduces the number of models from $2^{p + {p \choose 2}}$ to $\sum_{i = 0}^p {p \choose i} 2^{{i \choose 2}}$ or $ \sum_{i = 0}^p {p \choose i} 2^{p i - i (i+1)/2}$ for strong or weak heredity, respectively \citep{Chipman1996}. For ultra high-dimensional problems, two stage-approaches have been developed, see \cite{HaoFengZhang2016} and \cite{wang2019penalized}. However, these methods do not report uncertainties in model selection and parameter estimation, and rely on strong sparsity assumptions.

We are particularly motivated by studies of environmental health collecting data on mixtures of chemical exposures. These exposures can be moderately high-dimensional with high correlations within blocks of variables; for example, this can arise when an individual is exposed to a product having a mixture of chemicals and when chemical measurements consist of metabolites or breakdown products of a parent compound. There is a large public heath interest in studying E$\times$E, E$\times$G and G$\times$G interactions, with E = environmental exposures and G = genetic factors. However, current methods for quadratic regression are not ideal in these applications due to the level of correlation in the predictors, the fact that strong sparsity assumptions are not appropriate, and the need for uncertainty quantification. Regarding the issue of sparsity, some exposures are breakdown products of the same compound, so it is unlikely that only one exposure has an effect on the outcome. Also, it is statistically challenging to tell apart highly correlated covariates with limited data. For this reason, it is appealing given the data structure to select blocks of correlated exposures together instead of arbitrarily selecting one chemical in a group. 

To address these problems, one possibility is to use a Bayesian approach to inference in order to include prior information to reduce dimensionality while characterizing uncertainty through the posterior distribution. There is an immense literature on Bayesian methods for high-dimensional linear regression, including recent algorithms that can scale up to thousands of predictors \citep{bondell2012consistent}, \citep{rossell2017nonlocal}, \citep{johndrow2017scalable}, \citep{nishimura2018prior}. In addition some articles have explicitly focused on quadratic regression and interaction detection \citep{zhang2007bayesian}, \citep{cordell2009detecting}, \citep{mackay2014epistasis}. Bayes variable selection and shrinkage approaches will tend to have problems when predictors are highly correlated; this has motivated a literature on Bayesian latent factor regression \citep{lucas2006sparse}, \citep{Carvalho2008}. 

Latent factor regression incorporates shared latent variables in the predictor and response components. This provides dimensionality reduction in modeling of the covariance structure in the predictors and characterizing the impact of correlated groups of predictors on the response. Such approaches are closely related to principal components regression, but it tends to be easier to simultaneously incorporate shrinkage and uncertainty quantification within the Bayesian framework. In addition, within the Bayes latent factor regression paradigm, typical identifiability constraints such as orthogonality are not needed (see, for example \cite{Bhattacharya2013}). The main contribution of this article is to generalize Bayesian latent factor regression to accommodate interactions using an approach inspired by \cite{wang2019penalized}. This is accomplished by including pairwise interactions in the latent variables in the response component. We refer to the resulting framework as Factor analysis for INteractions (FIN). There is a rich literature on quadratic and nonlinear latent variable modeling, largely in psychometrics (refer, for example, to \cite{arminger1998bayesian}). However, to our knowledge, such approaches have not been used for inferences on interactions in regression problems. 

In \textit{Section 2} we describe the proposed FIN framework, including extensions for higher order interactions. In \textit{Section 3} we provide theory on model misspecification and consistency. \textit{Section 4} contains a simulation study. \textit{Section 5} illustrates the methods on NHANES data. Code is available at \url{https://github.com/fedfer/factor_interactions}. Proofs of \textit{Proposition 2} and \textit{Proposition 3} are included in the \textit{Supplementary Material}. 
\vspace*{-0.3cm}

\section{Model}
\vspace*{-0.3cm}
\subsection{Model and Properties}
Let $y_i$ denote a continuous health response for individual $i$, and $X_i = (x_{i1},\cdots,x_{ip})^T$ denote a vector of exposure measurements. We propose a latent factor joint model, which includes shared factors in both the predictor and response components while assuming conditional independence. We include interactions among latent variables in the response component. We also assume that, given the latent variables, the explanatory variables and the response are continuous and normally distributed. We assume that the data have been normalized prior to the analysis so that we omit the intercept. The model is as follows:
\begin{align}
& y_i = \eta_i^T \omega + \eta_i^T \Omega \eta_i +\epsilon_{y,i},  \quad \epsilon_{y,i} \sim N(0,\sigma^2) \nonumber,  \\ 
& X_i = \Lambda \eta_i+ \epsilon_i,   \quad  \epsilon_i \sim N_p(0,\Psi), \\
& \eta_i \sim N_k(0,I), \nonumber
\end{align}
where $\Psi = diag(\sigma_1^2,\cdots,\sigma^2_p)$. In a Bayesian fashion, we assume a prior for the parameters $\Theta = (\omega,\Omega,\Lambda,\Psi,\sigma^2)$ that will be specified in \textit{Section 2.2}. Model $(1)$ is equivalent to classical latent factor regression models; refer, for example, to \cite{Mike03}, except for the $\eta_i^T \Omega \eta_i$ term. Here, $\Omega$ is a $k \times k$ symmetric matrix inducing a quadratic latent variable regression that characterizes interactions among the latent variables.
 
The above formulation can be shown to induce a quadratic regression of $y$ on $X$. To build intuition consider the case in which $\sigma_j^2 = 0$ as done in \cite{Mike03} for the special case in which $\Omega = 0$. \textit{The many-to-one map} $X_i = \Lambda \eta_i$  has multiple generalized inverses $\eta_i = \Lambda^T X_i + b$ such that $\Lambda b = 0$. If we substitute in the regression equation, we obtain
\begin{align*}
 \E(y_i|X_i) & = (\Lambda^T X_i + b)^T \omega + (\Lambda^T X_i + b)^T \Omega (\Lambda^T X_i + b) = \\ 
 & = X_i^T \Lambda \omega + X_i^T \Lambda \Omega \Lambda^T X_i + g(b)
\end{align*}
The following proposition gives a similar result in the non deterministic case:
\begin{prop}
Under model (1), the following are true:
\vspace{-0.3cm}
\begin{itemize}
\item[(i)] $\E(y_i | X_i) = tr (\Omega V)+(\omega^T A) X_i + X_i^T (A^T \Omega A) X_i$,
\vspace{-0.3cm}
\item[(ii)] $Cov(y_i,X_i) = \Lambda \omega$,
\end{itemize}
\vspace{-0.3cm}
where $V = (\Lambda^T \Psi^{-1} \Lambda + I)^{-1}$ and $A = V \Lambda^T \Psi^{-1}  = (\Lambda^T \Psi^{-1} \Lambda + I)^{-1} \Lambda^T \Psi^{-1}$. 
\end{prop}
This shows that the induced regression of $y$ on $X$ from model (1) is indeed a quadratic regression. Let us define the induced main effects as $\beta_X = A^T \omega$ and the matrix containing the first order interactions as $\Omega_X = A^T \Omega A$. Notice that we could define $\Omega$ as a diagonal matrix and we would still estimate pairwise interactions between the regressors, further details are given in \textit{Sections 2.3} and \textit{2.5}.

In epidemiology studies, it is of interest to include interactions between chemical exposures and demographic covariates. The covariates are often binary variables, like \textit{race} or \textit{sex}, or continuous variables that are non-normally distributed, like \textit{age}. Hence, we do not want to assume a latent normal structure for the covariates. Letting $Z_i = (z_{i1}, \cdots, z_{iq})^T$ be a vector of covariates, we modify model $(1)$ to include a main effect for $Z_i$ and an interaction term between $Z_i$ and the latent factor $\eta_i$:
\begin{align}
& y_i = \eta_i^T \omega + \eta_i^T \Omega \eta_i + Z_i^T \alpha + \eta_i^T \Delta Z_i + \epsilon_{y,i},  \quad \epsilon_{y,i} \sim N(0,\sigma^2) \nonumber,  \\ 
& X_i = \Lambda \eta_i+ \epsilon_i,   \quad  \epsilon_i \sim N_p(0,\Psi), \\
& \eta_i \sim N_k(0,I), \nonumber
\end{align}
where $\Delta$ is a $k \times q$ matrix of interaction coefficients between the latent variables and the covariates, and $\alpha = (\alpha_1, \cdots, \alpha_q)$ are main effects for the covariates. Following \textit{Proposition 1} we have that
\begin{align*}
\mathbb{E}(\eta_i^T \Delta Z_i | X_i, Z_i ) =  \mathbb{E}(\eta_i^T  | X_i )\Delta Z_i =  X_i^T (A^T \Delta) Z_i ,
\end{align*}
where $(A^T \Delta)$ is a $p \times q$ matrix of pairwise interactions between exposures and covariates. In the sequel, we focus our development on model $(1)$ for ease in exposition, but all of the details can be easily modified to pertain to model $(2)$.

\subsection{Priors and MCMC Algorithm}

In this section we define the priors for $(\omega,\Omega,\Lambda,\Psi,\sigma^2)$, briefly describe the computational challenges given by model $(1)$ and summarize our Markov Chain Monte Carlo sampler in \textit{Algorithm 1}.  We choose an Inverse-Gamma distribution with parameters $(\frac{1}{2},\frac{1}{2})$ for $\sigma^2$ and $\sigma_j^2$ for $j = 1,\cdots,p$. The elements of $\omega$ and $\Omega$ are given independent Gaussian priors. For $\Lambda = \{ \lambda_{i,j} \}$, a typical choice to attain identifiability requires $\lambda_{i,j} = 0$ for $j > i$ and $\lambda_{j,j} > 0$ for $j = 1,\cdots,k$ \citep{geweke1996measuring}. However, some Bayesian applications, like covariance estimation \citep{Bhattacharya2013}, do not require identifiability of $\Lambda$. The same holds for inference on induced main effects and interactions for model $(1)$. Notice that model $(1)$ is invariant to rotations:
\begin{align*}
& y_i = \eta_i^T P P^T \omega + \eta_i^T PP^T\Omega P P^T \eta_i + \epsilon_{y,i},  \quad \epsilon_{y,i} \sim N(0,\sigma^2) \nonumber,  \\ 
& X_i = \Lambda P P^T \eta_i+ \epsilon_i,   \quad  \epsilon_i \sim N_p(0,\Psi),
\end{align*}
where $P$ is a $k \times k$ orthogonal matrix $P$ ($PP^T =I$). However, the induced main effects satisfy 
\begin{align*}
 \beta_X = \Psi^{-1} \Lambda  P ( P^T \Lambda^T  \Psi^{-1} \Lambda  P + P^T P)^{-1}  P^T \omega = \Psi^{-1} \Lambda ( \Lambda^T  \Psi^{-1} \Lambda  + I)^{-1} \omega.
\end{align*}
The same holds for induced interactions, showing that we do not need to impose identifiability constraints on $\Lambda$. We choose the Dirichlet-Laplace (DL) prior of \cite{Bhattacharya2015} row-wise, corresponding to
\begin{gather*}
\lambda_{j,h} | \phi_{jh}, \tau_j \sim DE(\phi_{jh} \tau_j) \ \ \ \ \ \ \ h = 1, \cdots, k  \\
\phi_j \sim Dir(a,\cdots,a) \ \ \ \ \ \ \tau_j \sim Gamma(k a , 1/2),
\end{gather*}
where $j = 1,\cdots,p$, $\phi_j = (\phi_{j1},\cdots,\phi_{jk})$, DE refers to the zero mean double-exponential or Laplace distribution, and $k$ is an upper bound on the number of factors, as the prior allows effective deletion of redundant factor loadings through row-wise shrinkage. 
The DL prior provides flexible shrinkage on the factor loadings matrix, generalizing the Bayesian Lasso \citep{park2008bayesian} to have a carefully chosen hierarchical structure on exposure-specific ($\tau_j$) and local ($\phi_{jh}$) scales. This induces a prior with concentration at zero, to strongly shrink small signals, and heavy-tails, to avoid over-shrinking large signals. The DL prior induces near sparsity row-wise in the matrix $\Lambda$, as it is reasonable to assume that each variable loads on few factors. 

In \textit{Section 2.4}, we describe how the above prior specification induces an appealing shrinkage prior on the main effects and interactions, and discuss hyperparameter choice. In practice, we recommend the rule of thumb that chooses $k$ such that $\frac{\sum_{j = 1}^k v_j}{\sum_{j = 1}^p v_j} > 0.9$, where $v_j$ is the $j^{th}$ largest singular value of the correlation matrix of $X$. \textit{Proposition 2} in \textit{Section 3} provides theoretical justification for this criterion. As an alternative to row-wise shrinkage, we could have instead used column-wise shrinkage as advocated in \cite{Bhattacharya2015} and \cite{legramanti2019bayesian}. Although such approaches can be effective in choosing the number of factors, we found in our simulations that they can lead to over-shrinkage of the estimated main effects and interactions.  


The inclusion of pairwise interactions among the factors in the regression of the outcome $y_i$ rules out using a simple data augmentation Gibbs sampler, as in \cite{Mike03}, \cite{Bhattacharya2013}. The log full conditional distribution for $\eta_i$ is:
\begin{align*}
& -\frac{1}{2} \Big[ \eta_i^T(\frac{\omega \omega^T}{\sigma^2_y}+\Lambda^T \Psi^{-1}\Lambda + I - 2 \frac{ \Omega Y_i }{\sigma_y^2})\eta_i  - 2 \eta_i ^T (\Lambda^T \Psi^{-1}X_i + \frac{\omega Y_i}{\sigma_y^2})\Big] - \\ 
& -\frac{1}{2} \Big[ \frac{2 \eta_i^T \omega \eta_i^T \Omega \eta_i}{\sigma_y^2} + \frac{(\eta_i^T \Omega \eta_i)^2}{\sigma_y^2}  \Big] + C,
\end{align*}
where $C$ is a normalizing constant. We update the factors $\eta_i$ using the Metropolis-Adjusted Langevin Algorithm (MALA) \citep{grenander1994representations}, \citep{roberts1996exponential}. Sampling the factors is the main computational bottleneck of our approach since we have to update $n$ vectors, each of dimension $k$. The overall MCMC algorithm and the MALA step are summarized in \textit{Algorithm 1}.

\begin{algorithm}[p]
{\footnotesize \caption{MCMC algorithm for sampling the parameters of model $(1)$}}
\begin{algorithmic} 
{\footnotesize
\vspace{0.05cm}
\STATE \textit{Step 1} Sample $\eta_i$, $i = 1, \cdots,n$ via Metropolis-Hastings using as a proposal distribution a $N\big(\eta_i + \hspace*{0.9cm}\frac{1}{2}\nabla_{\eta_i } log(\pi(\eta_i | \hbox{ --- }) \big), \epsilon I_k)$.
\STATE \textit{Step 2} Sample the main effects coefficients $\omega$ from a multivariate normal distribution:
\vspace{-0.2cm}
\begin{align*}
\pi(\omega |  \hbox{ --- }  ) \sim N \bigg( (\frac{\eta^T \eta}{\sigma^2} + I_n/100)^{-1} \eta \big(y - diag(\eta \Omega \eta)\big)/\sigma^2,(\frac{\eta^T \eta}{\sigma^2} + I_n/100)^{-1} \bigg)
\end{align*}\\[-0.13in]
\hspace*{0.9cm} where $\eta$ is the matrix with rows equal to $\eta_i$.
\STATE \textit{Step 3} Sample upper triangular part of $\Omega$, namely $\Omega^{\text{U}}$, from a multivariate normal distribution:
\vspace{-0.2cm}
\begin{align*}
\pi(\Omega^{\text{U}} |  \hbox{ --- }  ) \sim N \bigg( (\frac{\eta^{*T} \eta^*}{\sigma^2} + \frac{p(p + 1)}{2})^{-1} \eta^* \big(y - \eta \omega)\big)/\sigma^2,(\frac{\eta^{*T} \eta^*}{\sigma^2} + I_\frac{p(p + 1)}{2}/100)^{-1} \bigg)
\end{align*}\\[-0.13in]
\hspace*{0.9cm} where $\eta^*$ is a matrix containing the pairwise interactions of among the columns of $\eta$. Then set \hspace*{0.9cm} $\Omega = \frac{\Omega + \Omega^T}{2}$ 
\vspace{-0.2cm}
\STATE   \textit{Step 4} Sample $\sigma^{-2}$ from a Gamma distribution:
\vspace{-0.2cm}
\begin{align*}
\pi (\sigma^{-2} | \hbox{ --- } ) \sim Gamma \bigg( \frac{1 + n}{2}, \frac{1}{2} + \frac{1}{2} (y - \eta \omega - diag(\eta \Omega \eta^T))^T(y - \eta \omega - diag(\eta \Omega \eta^T)  \bigg)
\end{align*}\\[-0.13in]
\STATE  \textit{Step 5} Denote $\lambda_j$ the rows of $\Lambda$, for $j = 1,\cdots,p$. Sample $p$ conditionally independent posteriors:
\vspace{-0.2cm}
\begin{align*}
\pi (\lambda_j | \hbox{ --- } ) \sim N \bigg( (D_j^{-1} + \frac{\eta^T\eta}{\sigma_j^2})^{-1} \eta^T \sigma_j^{-2} X^{(j)},(D_j^{-1} + \frac{\eta^T\eta}{\sigma_j^2})^{-1} \bigg)
\end{align*}\\[-0.13in]
\hspace*{0.9cm} where $X^{(j)} $ is the $j^{th}$ column of the matrix $X$, $D_j = diag(\tau_j^2 \psi_{j1} \phi_{j 1},\cdots, \tau_j^2 \psi_{jk} \phi_{j k})$.
\STATE   \textit{Step 6} Sample $\psi_{jh}$ for $j = 1, \cdots, p$ and $h = 1 \cdots, k$ from independent Inverse Gaussian distributions:
\vspace{-0.2cm}
\begin{align*}
\pi (\psi_{jh} | \hbox{ --- } ) \sim InvGauss \bigg( \tau_j \phi_{jh} , 1 \bigg)
\end{align*}\\[-0.13in]
\STATE   \textit{Step 7} Sample $\tau_{j}$ for $j = 1, \cdots, p$  from independent Generalized \hspace*{0.9cm} Inverse Gaussian distributions:
\vspace{-0.2cm}
\begin{align*}
\pi (\tau_{j} | \hbox{ --- } ) \sim GInvGauss \bigg( 1-k, 1, 2 \sum_{h = 1}^k \frac{|\lambda{jh}|}{\phi_{jh}} \bigg)
\end{align*}\\[-0.43in]
\STATE   \textit{Step 8} In order to update $\phi_{jh}$ for $j = 1, \cdots, p$ and $h = 1 \cdots, k$, sample $T_{jh}$ from independent Generalized \hspace*{0.9cm} Inverse Gaussian distributions:
\vspace{-0.2cm}
\begin{align*}
\pi (T_{jh} | \hbox{ --- } ) \sim GInvGauss \bigg( a- 1 , 1 , 2 |\lambda_{jh}| \bigg)
\end{align*}\\[-0.13in]
\hspace*{0.9cm} Then set $\phi_{jh} = \frac{T_{jh}}{\sum_{h = 1}^k T_{jh} }$
\STATE   \textit{Step 9} Sample $\sigma_j^{-2}$ for $j = 1,\cdots, p$ from conditionally independent gamma distributions
\vspace{-0.2cm}
\begin{align*}
\pi (\sigma_j^{-2} | \hbox{ --- } ) \sim Gamma \bigg( \frac{1 + n}{2}, \frac{1}{2} + \frac{1}{2} \sum_{i = 1}^n (X_{ij} - \lambda_j^T \eta_i )\bigg)
\end{align*}\\[-0.13in]
\vspace{-0.1cm}
}
\end{algorithmic}
\end{algorithm}

\subsection{Higher Order Interactions}

FIN can be generalized to allow for higher order interactions. In particular,  we can obtain estimates for the interaction coefficients up to the $Q^{th}$ order with the following model:
\begin{align}
\E(y_i| \eta_i) = \sum_{h= 1}^k \omega_h^{(1)} \eta_{i h} + \sum_{h = 1}^k \omega_h^{(2)} \eta_{i h} ^2 + \cdots + \sum_{h= 1}^k \omega_h^{(Q)} \eta_{i h} ^Q,
\end{align}
which is a polynomial regression in the latent variables. We do not include interactions between the factors, so that the number of parameters to be estimated is $Q  k$. When $Q=2$, this model is equivalent to $\Omega$ being a diagonal matrix. Recall that $\eta_i | X_i \sim N(A X,  V)$, where $A$ and $V$ are defined in \textit{Proposition 1}. Since we do not include interactions among the factors, let us just focus on the marginal distribution of the $j^{th}$ factor, i.e $\eta_{i h} | X_i \sim N(\mu_h, \sigma_h^2)$ where $\mu_h = \sum_{j = 1} ^ p a_{hj}X_{i j}$ and $\sigma_h^2 = V_{hh}$. Below we provide an expression for $E( \sum_{q = 1}^Q \omega_h^{(q)} \eta^q_{ih} | X )$, which can be calculated using non-central moments of a Normal distribution, see \cite{winkelbauer2012} for a reference.
\begin{align*}
\E( \sum_{q = 1}^Q \omega_j^{(q)} \eta^q_j | X ) = &
\sum_{f = 1}^{\lfloor \frac{Q+1}{2} \rfloor }   \sum_{q = f}^{\lfloor \frac{Q+1}{2} \rfloor}  \omega_h^{(2q-1)} \sigma_h^{2 q -2f}   b_{qf}^{o}  \sum_{k_+= 2 f -1} \binom{2f -1}{k_1 \cdots k_p} \prod_{j = 1}^p (a_{h j} X_j)^{k_j} + \\  & \sum_{f = 0}^{\lceil \frac{Q+1}{2} \rceil }   \sum_{q = f \vee 1}^{\lceil\frac{Q+1}{2} \rceil}  \omega_h^{(2q)} \sigma_h^{2 q  -2f }     b_{qf}^{e}    \sum_{k_+ = 2f} \binom{2f}{k_1 \cdots k_p} \prod_{j = 1}^p (a_{hj} X_j)^{k_j},
\end{align*}
where $b_{qf}^{o} = \frac{(2 q -1 )!}{(2 f -1)! (q-f)! 2^{q-f}}   $, $b_{qf}^{e} =  \frac{(2 q  )!}{(2 f)! (q-f)! 2^{q-f}} $ and $k_+ = \sum_{j = 1}^p k_h$. We just need to sum up over the index $h$ in $(3)$ and we can read out the expressions for the intercept, main effects and interactions up to the $Q^{th}$ order. In particular, we have that the intercept is equal to $\sum_{h = 1}^k \sum_{q = 1}^{\lceil \frac{Q-1}{2} \rceil } \omega_{h}^{(2q)} V^{2 q }_{hh} b_{q0}^e $. When $Q = 2$ this reduces to $\sum_{h = 1}^k \omega_{h}^{(2)} V^{2  }_{hh}  = tr(\Omega V) $, where $\Omega = diag(\omega_1^{(2)},\cdots,\omega_k^{(2)})$. The expression for the main effects coefficients on $X_j$ is $ \sum_{h = 1}^k \sum_{q = 1}^{\lfloor \frac{Q+1}{2} \rfloor } \omega_{h}^{(2q-1)} V^{2 q -1 }_{hh} b_{q1}^o a_{hj} $. When $Q = 2$ this becomes $ \sum_{h= 1}^k \omega_{h}^{(1)} a_{hj}$, hence $\beta_X = A^T \omega$.  Similarly the expression for the interaction between $X_j$ and $X_l$ is equal to $ \sum_{h = 1}^k \sum_{q = 1}^{\lceil \frac{Q-1}{2} \rceil  } 2 \omega_{h}^{(2q)} V^{2 q }_{hh} b_{q1}^e a_{hj}  a_{hl}$ and when $Q = 2$ we have $\sum_{h = 1}^k 2 \omega_{h}^{(2)} a_{hj}  a_{hl}$ which is equal to $2 [A^T \Omega A]_{(j,l)}$. 

In general, if we are interested in the $q^{th}$ order interactions, we can find the expression on the top summation for $f = \frac{q+1}{2}$ when $q$ is odd and on the bottom summation for  $f = \frac{q}{2}$ when $q$ is even. Finally notice that with $Q k$ parameters we manage to estimate $\sum_{q =  0}^Q {p \choose q}$ parameters thanks to the low dimensional factor structure in the covariates.

\subsection{Induced Priors}

In this section, we show the behavior of the induced priors on the main effects and pairwise interaction coefficients under model $(1)$ using simulated examples, and we show the induced grouping of coefficients when we have prior information on the covariance structure of $X$. We endow $\omega$ with a normal prior having zero mean  and covariance equal to $\Xi$, where $\Xi$ is a diagonal matrix. Then, conditional on $\Lambda$ and $\Psi$, the induced prior on $\beta_X$ is also Normal with mean $0$ and covariance equal to $A^T A$. Recall from \textit{Proposition 1} that the induced main effect coefficients are equal to $\beta_X^T = \omega^T (\Lambda^T \Psi^{-1} \Lambda)^{-1} \Lambda^T \Psi^{-1} $. This expression is equivalent to \cite{Mike03} and we can similarly characterize the limiting case of $\Psi \rightarrow 0$, i.e. when the factors explain all of the variability in the matrix of regressors $X_i$. Let $\Psi = s I$ and $s \rightarrow 0$, together with enforcing $\Lambda$ to be orthogonal, we have that $\beta_X  = \Lambda \omega$. It follows that $\beta_X$ has the \textit{generalised singular g-prior} (or \textit{gsg}-prior) distribution defined by \cite{Mike03}, whose density is proportional to $ exp(-\frac{1}{2} \beta_X^T \Lambda^T \Xi^{-1} \Lambda \beta)$.

Now, consider the extension presented in the previous section, where we include powers of the factors in the regression of $y_i$. In \textit{Figure 1}, we show the induced marginal priors for main effects, pairwise interactions and $3^{rd}$ order interactions when $p = 20$ and $k = 5,10$ when $\omega$ and $\Omega$ are given $N(0,1)$ priors element-wise. Increasing (or decreasing) the variance of the priors on $\omega$ and $\Omega$ will directly increase (or decrease) the variance of the induced main effects and pairwise interactions, as $\beta_X$ and $\Omega_X$ are linear functions of $\omega$ and $\Omega$ respectively. For a fixed $k$, there is increasing shrinkage towards zero with higher orders of interaction. However, we avoid assuming exact sparsity corresponding to zero values of the coefficients, a standard assumption of other methods. Although most of the mass is concentrated around zero, the distributions have heavy tails. We can indeed notice that the form of the priors resembles a mixture of two normal distributions with different variances, and that we place a higher mixture weight on the normal distribution concentrated around zero as we increase the order of interactions. This is because higher order interactions contain products of the elements of $A$, previously defined in \textit{Proposition 1}, and the elements of $A$ are affected by the DL prior shrinkage, since $A$ is a function of $\Lambda$. Also, notice that the priors have higher variance as we increase the number of latent factors $k$.

\begin{figure}[htbp]
\begin{center}
  \includegraphics[width=\linewidth, height = 7cm]{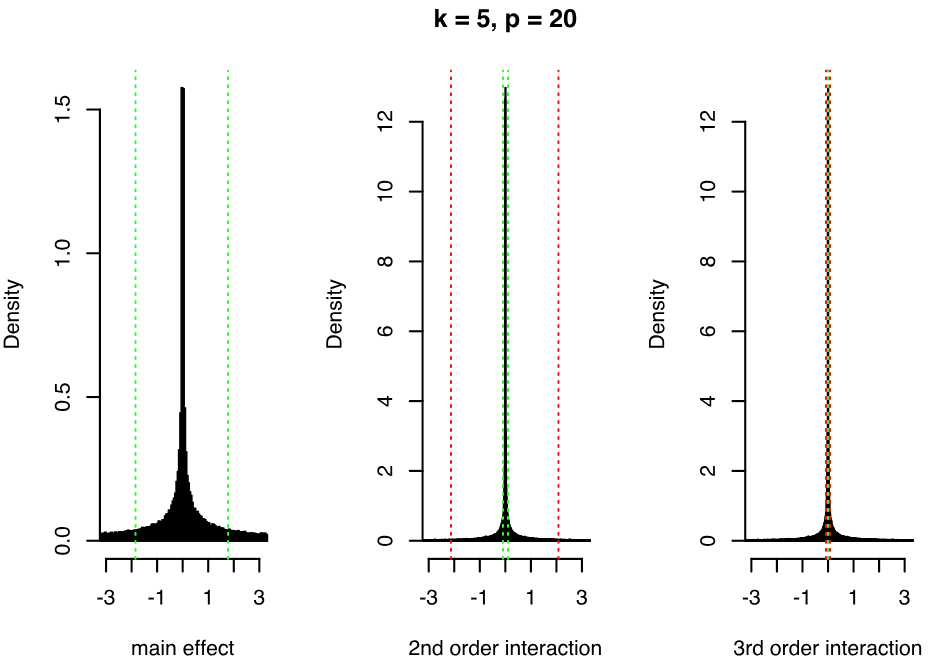}
   \includegraphics[width=\linewidth, height = 7cm]{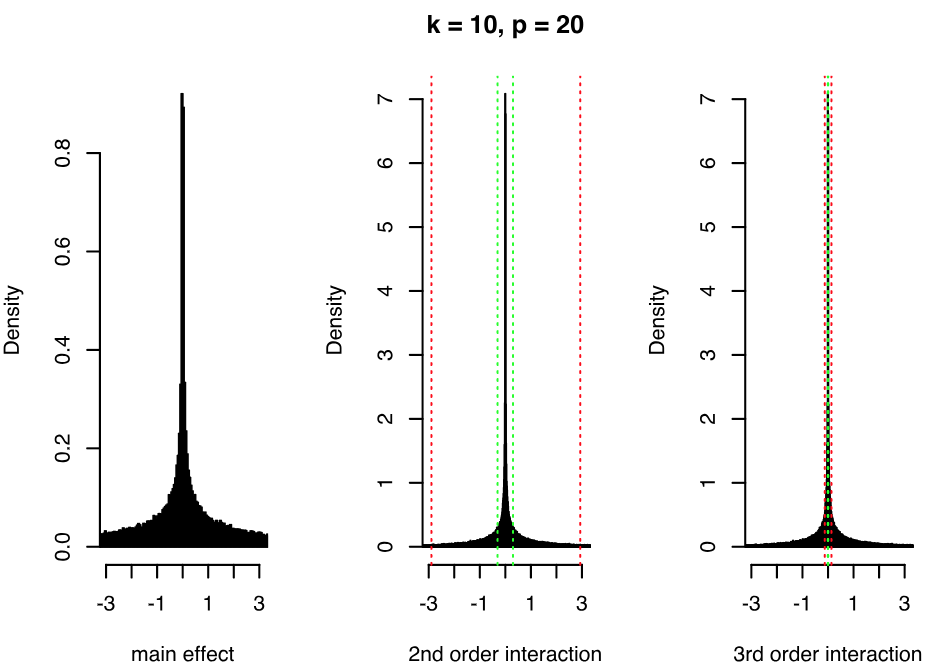}
  \label{fig:boat1}
\end{center}
  \caption{Induced priors on main effects, pairwise interactions and $3^{rd}$ order interactions for $p = 20$ and $k = 5,10$. The green lines corresponds to $0.25$ and $0.75$ quartiles and the red lines to the $0.05$ and $0.95$.}
\end{figure}

In environmental epidemiology, it is common to have prior knowledge of groups of exposures that are highly correlated and it is natural to include such information in the specification of $\Lambda$. One possibility is to impose a block sparsity structure in which each group of chemicals is restricted to load on the same factor. Then, cross group dependence is allowed including additional factors and endowing the factor loadings with a DL prior. Suppose that the variables in $X$ can be divided in $l$ groups: $S_1, S_2, \ldots S_l$ of dimensions $p_1,p_2,\ldots,p_l$, where $l < k$ and $p = \sum_{r = 1}^l p_r$. Letting $\Lambda = [\Lambda^B \Lambda']$, where $\Lambda^B$ is $p \times l$, we can assign a block sparsity structure to $\Lambda^B$:
\begin{align*}
    & \lambda^B_{p_1 +1, 1} = \ldots = \lambda^B_{p, 1} = 0 \\ 
    & \lambda^B_{1, 2} = \ldots = \lambda^B_{p_1, 2} = \lambda^B_{p_1 + p_2 + 1, 2} = \ldots =  \lambda^B_{p, 2} = 0 \\
   & \cdots \\
    & \lambda^B_{1, l} = \ldots = \lambda^B_{p - p_l, l} = 0 
\end{align*}
In the \textit{Supplementary Material} we show the effect of the block sparsity structure on the a priori induced groupings of main effects and interactions when $l = k$, so that $\Lambda = \Lambda^B$.

\section{Properties of the Model}
In this section we prove that the posterior distribution of $\Theta = (\omega,\Omega,\sigma^2,\Lambda,\Psi)$ is weakly consistent for a broad set of models. Let $KL(\Theta_0,\Theta)$ denote the Kullback-Leibler divergence between $p(X,y |	\Theta_0) $ and $p(X,y|\Theta)$, where
\begin{align*}
p(X,y|\Theta_0) =  \int p(X| \Lambda_0,\Psi_0,\eta)p(y|\omega_0,\Omega_0,\sigma_0^2,\eta) p(\eta) d\eta .
\end{align*}
We will assume that $p(X,y|\Theta_0) $ represents the true data-generating model. This assumption is not as restrictive as it may initially seem. The model is flexible enough to always characterize and model quadratic regression in the response component, while accurately approximating any covariance structure in the predictor component. In fact it always holds that:
\begin{align*}
& \mathbb{E}(y_i| X_i) = \beta_0 X_i + X_i \Omega_0 X_i , \\
& X_i \sim N(0,\Lambda_0 \Lambda_0 ^T + \Psi_0), 
\end{align*}
where $\beta_0$ and $\Omega_0 $ are functions of $\Theta_0$ as in \textit{Proposition 1}, and the true number of factors is $k_0$. When $k_0 = p$, we can write any covariance matrix as $\Lambda_0 \Lambda_0^T + \Psi_0$.
We take an ``over-fitted'' factor modeling approach, related to \cite{Bhattacharya2013}, \cite{rousseau2011asymptotic}, and choose $k$ to correspond to an upper bound on the number of factors. In practice, we recommend the rule of thumb that chooses $k$ such that $\frac{\sum_{j = 1}^k v_j}{\sum_{j = 1}^p v_j} > 0.9$, where $v_j$ is the $j^{th}$ largest singular value of the correlation matrix of $X$. We have found this choice to have good performance in a wide variety of simulation cases. However, there is nonetheless a potential concern that $k$ may be less than $k_0$ in some cases. \textit{Proposition 2} quantifies the distance in terms of Kullback-Leibler divergence between the true data generating model and the likelihood under model miss-specification as $n$ approaches infinity.
\begin{prop}
Fix $\Lambda_0,\Psi_0 = s_0 I_p$, $k_0$,  and assume that $k<k_0$. As $n$ increases the posterior distribution of $\Lambda$ and $\Psi = s I_p$ concentrates around $\Lambda^*$ and $\Psi^*$, satisfying:
\begin{align*}
KL((\Lambda_0,\Psi_0);(\Lambda^*,\Psi^*)) \le \sum_{j = k + 1}^{k_0} \frac{ v_j}{s_0 } ,
\end{align*}
where $v_j$ is the $j^{th}$ largest singular value of $\Lambda_0 \Lambda_0 ^T$.
\end{prop}
Unsurprisingly, the bound of \textit{Proposition 2} resembles the Eckart-Young theorem for low-rank approximation based on the Singular Value Decomposition of a matrix. The Eckart-Young theorem states that the rank $k$ approximation $\hat{\Omega}$ of a matrix $\Omega$ minimizing the Frobenoius norm is such that $||\hat{\Omega} - \Omega||_F  = \sqrt{ \sum_{j = k+1}^p v_j^2} $. In a similar fashion as Principal Component Analysis and Factor Analysis, we can inspect the singular values of the correlation matrix of the regressors in order to choose the number of factors to include in the model, and thanks to \textit{Proposition 2} we know how far the posterior distribution will concentrate from the truth. 

The next proposition provides a sufficient condition in order to achieve posterior consistency when $k \ge k_0$. Notice that we achieve posterior consistency on the induced main effects and pairwise interactions.
\begin{prop}
Fix $\Theta_0 = (\omega_0,\Omega_0,\sigma_{0}^2,\Lambda_0,\Psi_0,k_0)$. Whenever $k \ge k_0$, for any $\delta > 0$ there exists an $\epsilon > 0$ such that:
\begin{align*}
\{\Theta: d_{\infty}(\Theta_0,\Theta)< \delta \} \subset \{\Theta: KL(\Theta_0,\Theta)< \epsilon \}
\end{align*}
where $d_\infty$ is the sup-norm.
\end{prop}
One can easily define a prior on $\Theta$ such that it places positive probability in any small neighborhood of $\Theta_0$, according to $d_\infty$. The prior defined in \textit{Section 2.2} satisfies this condition. Consequently, the posterior distribution of $\Theta$ is weakly consistent due to \cite{schwartz1965bayes}. The proofs of \textit{Proposition 2} and \textit{Proposition 3} can be found in the \textit{Supplementary Material}.

\section{Simulation Experiments}

In this section we compare the performance of our FIN method with four other approaches: PIE \citep{wang2019penalized}, RAMP \citep{HaoFengZhang2016}, Family \citep{Haris2018} and HierNet \citep{Bien2013}. These methods are designed for inference on interactions in moderate to high dimensional settings. We generate $25$ and $50$ covariates in three ways: 
\begin{align*}
&  X_i \sim N_p(0,\Lambda \Lambda^T +I_p), \quad \lambda_{i,j} \sim N(0,1), \quad && \text{(factor)} \\
&    X_i \sim N_p(0,W), \quad [W]_{i,j} = 0.8 |i - j|, \quad && \text{(linear)}\\
&    X_i \sim N_p(0,I_p). \quad && \text{(independent)}
\end{align*}
In the factor scenario we set the true number of factors equal to $7$ for $p=25$ and equal to $17$ when $p = 50$. FIN achieved similar performance when we chose a smaller number of latent factors. The average absolute correlation in the covariates is between $0.25$ and $0.3$ for the factor and linear scenarios when $p = 25$. These two simulation scenarios are the most similar to the environmental epidemiology data analysis in \textit{Section 5}. The complexity gains of FIN with respect to a Bayesian linear models with interactions is analyzed in the \textit{Supplementary Material}.

For each scenario, we generate the continuous outcome according to a linear regression with pairwise interactions:
\begin{align*}
y_i = X_i^T \beta_0  + X_i^T \Omega_0 X_i + \epsilon_i,
\end{align*}
where half of the main effects are different from zero and $\epsilon_i \sim N(0,1)$ for $i = 1,\cdots,500$. We distinguish between a sparse matrix of pairwise interactions $\Omega_0$, with only $5\%$ non-zero interactions, or dense, where $20\%$ of the elements are different from zero. 

For each value of $p$ we have six simulation scenarios: factor, linear or independent combined with sparse or dense pairwise interactions. We generate  the non-zero main effects and interaction coefficients from a Uniform distribution in the interval $(-1,-0.5) \bigcup (0.5,1)$ such that the regression equation follows the strong heredity constraint. Strong heredity allows an interaction between two variables to be included in the model only if the main effects are. This is done to favor RAMP, Family and HierNet, which assume the heredity condition. We repeat the simulations $50$ times and evaluate the performance on a test dataset of $500$ units computing predictive mean square error, mean square error for main effects, Frobenious norm (FR) for interaction effects, and percentage of true positives (TP) and true negatives (TN) for main effects and interactions. The percentage of TP and TN main effects is defined as follows:
\begin{align*}
& \text{TP}(\text{main effects}) = \frac{1}{p} \sum_{j = 1}^p \mathbbm{1}(\hat{\beta_j} \neq 0 ,\beta_{0j} \neq 0 , sign(\hat{\beta_j}) = sign(\beta_{0j}) ) \\
& \text{TN}(\text{main effects}) = \frac{1}{p} \sum_{j = 1}^p \mathbbm{1}(\hat{\beta}_j = 0,\beta_{0j} = 0   ),
\end{align*} 
where $\hat{\beta_j}$ is the estimated main effect for feature $j$ and $\beta_{0j}$ is the true coefficient. FIN is the only method reporting uncertainty quantification and we set $\hat{\beta_j} = 0$ whenever zero is included in the $95 \%$ credible interval. We equivalently define the percentage of true positive and true negative interactions. 

The MCMC algorithm was run for $5000$ iterations with a burn-in of $4000$. We observed good mixing. In particular, the Effective Sample Size (ESS) was always greater than $900$ across our simulations, both for main effects and interactions. We set the hyperparameter $a$ of the Dirichlet-Laplace prior equal to $1/2$. We obtained similar results for $a$ in the interval $[1/p,p]$. The results are summarized in \textit{Table 1-2} for $p = 25$ and in \textit{Table 1-2} of the \textit{Supplementary Material} for $p = 50$. Across all the simulations, we chose $k$ such that $\frac{\sum_{j = 1}^k v_j}{\sum_{j = 1}^p v_j} > 0.9$. 

In the factor scenario, FIN outperforms the other methods in predictive performance and estimation of main effects and interactions, whereas the rate recovery of true main effects and interactions is comparable to HierNet and PIE with sparse $\Omega_0$ and outperforms the other methods when $\Omega_0$ is dense. The latter scenario is the most challenging with respect to selection of main effects and interactions. Most of the other methods either select or shrink to zero all the effects. 
In the linear scenario, FIN also shows the best performance together with PIE and Hiernet. Despite the model misspecification with independent covariates, FIN has a comparable predictive performance with respect to the other methods, which do not take into account correlation structure in the covariates. The $95\%$ predictive intervals provided by FIN contained the true value of $y_i$ on average approximately $95\%$ of the time in the factor scenario, $89\%$ for the linear scenario, and $79\%$ for the independent scenario. The average bias in the posterior predictive mean is negligible in each simulation scenario.

The optimization method performed by HierNet \citep{Bien2013} tends to favor interactions only in presence of large component main effects, and in doing so overshrinks interactions estimates, especially in the \textit{dense scenario}. Penalized regression techniques PIE \citep{wang2019penalized} and RAMP \citep{HaoFengZhang2016} tend to over-shrink coefficient estimates and select too few predictors, particularly in the dense scenario. On the other hand, FAMILY \citep{Haris2018} performs a relaxed version of the penalized algorithm by refitting an unpenalized least squares model, which results in a high false positive rate of main effects. We also considered different signal-to-noise ratios with $\epsilon_i \sim N(0,\frac{1}{4})$ and $\epsilon_i \sim N(0,4)$. The results are very similar to the results we have presented; hence, we omit them.

\begin{table}[!p] \centering 
  \label{} 
   {\footnotesize
 {\renewcommand{\arraystretch}{0.7}
\begin{tabular}{@{\extracolsep{5pt}} c| cccccc} 
\\[-1.8ex]\hline 
\hline \\[-1.8ex] 
 & & HierNet & FAMILY & PIE & RAMP & FIN \\ 
\hline \\[-1.8ex] 
& test error & $1.974$ & $16.689$ & $7.067$ & $64.717$ & $1$ \\ 
& FR & $1.361$ & $1.013$ & $1.418$ & $1.620$ & $1$ \\ 
factor & main MSE & $1.167$ & $1.062$ & $1.807$ & $4.225$ & $1$ \\ 
\cline{2-7}
& TP main & $0.920$ & $0.988$ & $0.155$ & $0.270$ & $0.753$ \\ 
& TN main & $0.067$ & $0.007$ & $0.921$ & $0.773$ & $0.475$ \\ 
& TP int & $0.151$ & $0.807$ & $0.105$ & $0.037$ & $0.699$ \\ 
& TN int & $0.889$ & $0.233$ & $0.929$ & $0.962$ & $0.387$ \\ 
\hline 
\hline   \\[-1.8ex] 
& test error & $1$ & $2.662$ & $1.688$ & $6.309$ & $1.565$ \\ 
& FR & $1$ & $1.049$ & $1.075$ & $1.289$ & $1.016$ \\ 
linear & main MSE & $2.421$ & $1$ & $1.766$ & $4.259$ & $1.263$ \\
\cline{2-7}
& TP main & $1$ & $0.996$ & $0.177$ & $0.301$ & $0.572$ \\ 
& TN main & $0.002$ & $0.005$ & $0.904$ & $0.805$ & $0.718$ \\ 
& TP int & $0.532$ & $0.818$ & $0.280$ & $0.028$ & $0.635$ \\ 
& TN int & $0.849$ & $0.278$ & $0.887$ & $0.968$ & $0.570$ \\
\hline 
\hline   \\[-1.8ex] 
& test error & $1$ & $6.150$ & $2.759$ & $10.729$ & $7.128$ \\ 
& FR & $1.175$ & $1.548$ & $1$ & $2.042$ & $1.654$ \\ 
independent & main MSE & $1$ & $1.529$ & $1.756$ & $2.446$ & $2.031$ \\ 
\cline{2-7}
& TP main & $1$ & $1$ & $0.241$ & $0.074$ & $0.302$ \\ 
& TN main & $0$ & $0.002$ & $0.930$ & $0.985$ & $0.888$ \\ 
& TP int & $0.989$ & $0.952$ & $0.641$ & $0.005$ & $0.412$ \\ 
& TN int & $0.937$ & $0.414$ & $0.908$ & $1.000$ & $0.914$ \\ 
\hline 
\hline   \\[-1.8ex] 
\end{tabular} 
}
}
 {\footnotesize \caption{Results from simulation study with $p=25$ and \textit{dense} $\Omega_0$ in the three scenarios: factor, linear and independent for $n = 500$. We computed test error, Frobenious norm, MSE for main effects, percentage of true positives and true negatives for main effects and interactions for Hiernet, Family, PIE, RAMP and FIN model with $a = 0.5$ across $50$ simulations. Test error, FR, and main MSE are presented as ratios compared to the best performing model.}} 
\end{table}

\begin{table}[!p] \centering 
  \label{} 
   {\footnotesize
 {\renewcommand{\arraystretch}{0.7}
\begin{tabular}{@{\extracolsep{5pt}} c| cccccc} 
\\[-1.8ex]\hline 
\hline \\[-1.8ex] 
 & & HierNet & FAMILY & PIE & RAMP & FIN \\ 
\hline \\[-1.8ex] 
& test error & $1.284$ & $5.274$ & $1.206$ & $4.225$ & $1$ \\ 
& FR & $1.189$ & $1.259$ & $1$ & $2.157$ & $1.284$ \\ 
factor & main MSE & $3.430$ & $1.560$ & $1$ & $1.590$ & $1.312$ \\ 
\cline{2-7}
& TP main & $0.667$ & $0.823$ & $0.698$ & $0.583$ & $0.812$ \\ 
& TN main & $0.445$ & $0.259$ & $0.863$ & $0.834$ & $0.716$ \\ 
& TP int & $0.514$ & $0.839$ & $0.562$ & $0.031$ & $0.448$ \\ 
& TN int & $0.959$ & $0.580$ & $0.974$ & $0.965$ & $0.941$ \\
\hline 
\hline   \\[-1.8ex] 
& test error & $1.199$ & $5.271$ & $1$ & $5.060$ & $1.486$ \\ 
& FR & $3.889$ & $6.859$ & $1$ & $7.916$ & $5.370$ \\ 
linear & main MSE & $1$ & $3.563$ & $1.387$ & $1.392$ & $1.726$ \\ 
\cline{2-7}
& TP main  & $1$ & $0.845$ & $0.857$ & $0.952$ & $0.976$ \\ 
& TN main & $0.484$ & $0.272$ & $0.845$ & $0.815$ & $0.807$ \\ 
& TP int & $0.970$ & $0.887$ & $0.964$ & $0.077$ & $0.917$ \\ 
& TN int & $0.970$ & $0.645$ & $0.987$ & $0.975$ & $0.894$ \\ 
\hline 
\hline   \\[-1.8ex] 
& test error & $1.425$ & $9.685$ & $1$ & $12.746$ & $3.438$ \\ 
& FR & $12.956$ & $18.036$ & $1$ & $21.604$ & $9.607$ \\
independent & main MSE & $1$ & $6.082$ & $3.056$ & $4.326$ & $3.055$ \\ 
\cline{2-7}
& TP main & $1$ & $0.830$ & $0.860$ & $0.630$ & $0.900$ \\ 
& TN main & $0.418$ & $0.585$ & $0.847$ & $0.915$ & $0.898$ \\ 
& TP int & $1$ & $0.852$ & $1$ & $0.071$ & $0.921$ \\ 
& TN int & $0.993$ & $0.868$ & $0.990$ & $0.995$ & $0.957$ \\ 
\hline 
\hline   \\[-1.8ex] 
\end{tabular} 
}
}
 {\footnotesize \caption{Results from simulation study with $p=25$ and \textit{sparse} $\Omega_0$ in the three scenarios: factor, linear and independent for $n = 500$. We computed test error, Frobenious norm, MSE for main effects, percentage of true positives and true negatives for main effects and interactions for Hiernet, Family, PIE, RAMP and FIN model with $a = 0.5$ across $50$ simulations. Test error, FR, and main MSE are presented as ratios compared to the best performing model.}} 
\end{table}

\section{Environmental Epidemiology Application}
 
The goal of our analysis is to assess the effect of ten phthalate metabolites, four perfluoroalkyl (pfas) metabolites and fourteen metals on body mass index (BMI). Phthalates are mainly used as plasticizers and can be found in toys, detergents, food packaging, and soaps. They have previously been associated with increased BMI \citep{hatch2008association} and waist circumference (WC) \citep{stahlhut2007concentrations}. There is a growing health concern for the association of phthalates \citep{kim2014phthalate}, \citep{zhang2014age} and pfas metabolites \citep{braun2017early} with childhood obesity. Metals have already been associated with an increase in waist circumference and BMI, see \cite{padilla2010examination} and \cite{shao2017association}, using data from the National Health and Nutrition Examination Survey (NHANES). 

We also consider data from NHANES, using data from the years 2015 and 2016. We select a subsample of $7602$ individuals for which the measurement of BMI is not missing, though FIN can easily accommodate missing outcomes through adding an imputation step to the MCMC algorithm. \textit{Figure 3} contains a plot of the correlation between exposures. Several pairwise correlations are missing, as for example between pfas and most metals, because some chemicals are only measured within subsamples of the data. The average absolute correlation between the $28$ exposures is around $0.28$, similarly to the \textit{factor} and \textit{linear} simulation scenarios presented in \textit{Section 4}. We also include in the analysis cholesterol, creatinine, race, sex, education and age. We apply the $\text{log}_{10}$ transformation to the chemicals, cholesterol and creatinine. Histograms of the chemical measurements can be found in \textit{Figure 1} of the \textit{Supplementary Material}. We also apply the $\text{log}_{10}$ transformation to BMI in order to make its distribution closer to normality, which is the assumed marginal distribution in our model. The log-transformation is commonly applied in environmental epidemiology in order to reduce the influence of outliers and has been employed in several studies using NHANES data \citep{nagelkerke2006body}, \citep{lynch2010objectively}, \citep{buman2013reallocating}. We leave these transformations implicit for the remainder of the section. 

We assume a latent normal structure for the chemicals, which are included in the matrix $X$, and use the other variables as covariates, which are included in the matrix $Z$. We estimate a quadratic regression according to model $(2)$. We specify independent Gaussian priors for elements of $\alpha$ and $\Delta$. \textit{Algorithm 1} can be easily adapted for model $(2)$. The matrix $X$ has $60\%$ missing data and \textit{Figure 2} of the \textit{Supplementary material} contains a plot of the missingness pattern. Since we are modeling the chemical measurements, we can simply add a sampling step to the MCMC algorithm to sample the missing data according to $(2)$. Similarly, $0.4\%$ of chemicals have been recorded under the limit of detection (LOD). In order to be coherent with our model we can sample these observations as:
\begin{align*}
    X_{ij} | X_{ij} \in [-\infty, \text{log}_{10}(\text{LOD}_j)] \sim TN(\eta_i ^T \lambda_j, \sigma_j^2, -\infty, \text{log}_{10}(\text{LOD}_j))
\end{align*}
where $\text{LOD}_j$ is the limit of detection for exposure $j$ and $TN(\mu,\sigma^2, a, b)$ is a truncated normal distribution with mean $\mu$, variance $\sigma^2$ and support in $[a,b]$. We imputed the missing data using MICE \citep{white2011multiple} to compute the correlation matrix of chemicals. We noticed from the Eigendecomposition of the correlation matrix that the first $13$ eigenvectors explain more than $90\%$ of the total variability; hence we set the number of factors equal to $13$. 

\textit{Figure 2} on the right shows the posterior mean of the matrix of factor loadings $\Lambda$, before and after applying the MatchAlign algorithm of \cite{Evan2019}, which resolves rotational ambiguity and column label switching for the posterior samples of $\Lambda$. The matrix of factor loadings reflects the correlation structure of the chemicals. We can distinguish three families of chemicals: metals collected from urine, pfas and phathalates. The pfas chemicals load mostly on the $1^{st}$ factor, the metals from urine on the $8^{th}$ factor together with the phthalates, which is expected since there is high correlation between the two groups of chemicals. Finally, a group of highly correlated phthalates loads on the $13^{th}$ factor.

We also estimated a regression with pairwise interactions using the methods PIE, RAMP, Family and HierNet introduced in \textit{Section 4}. These methods do not directly deal with missing data, so we imputed the missing data using MICE \citep{white2011multiple}. \textit{Figure 3} shows the estimated main effects of the chemicals. The signs of the coefficients are generally consistent across different methods. 

\textit{Figure 4} shows the posterior mean of the matrix of chemical interactions and of the matrix $A^T\Delta$ of pairwise interactions between exposures and covariates. As expected, we estimate a ``dense" matrix of interactions. This is due to exposures being breakdown products of the same compound and high correlation between chemicals belonging to the same family. For example the correlation between the pfas metabolites is equal to $0.7$, with only $1977$ observations containing complete measurements.
Interactions between highly correlated pfas metabolites have been observed in animal studies \citep{wolf2014evaluating}, \citep{ding2013combined}. Linear \citep{henn2011associations}, \citep{lin2013utero} and nonlinear interactions \citep{valeri2017joint} between metals have been associated with neurodevelopment. Interactions between phathalates and other chemicals have been related to human semen quality \citep{hauser2005evidence}.
Finally, we estimate several interactions between chemicals and age, cholesterol and creatinine, which are usually expected in environmental epidemiology applications \citep{barr2004urinary}. 
The code for reproducing the analysis is available at \url{https://github.com/fedfer/factor_interactions}.

\begin{figure}[htbp]
\centering
\includegraphics[width=\linewidth, height= 6.5cm]{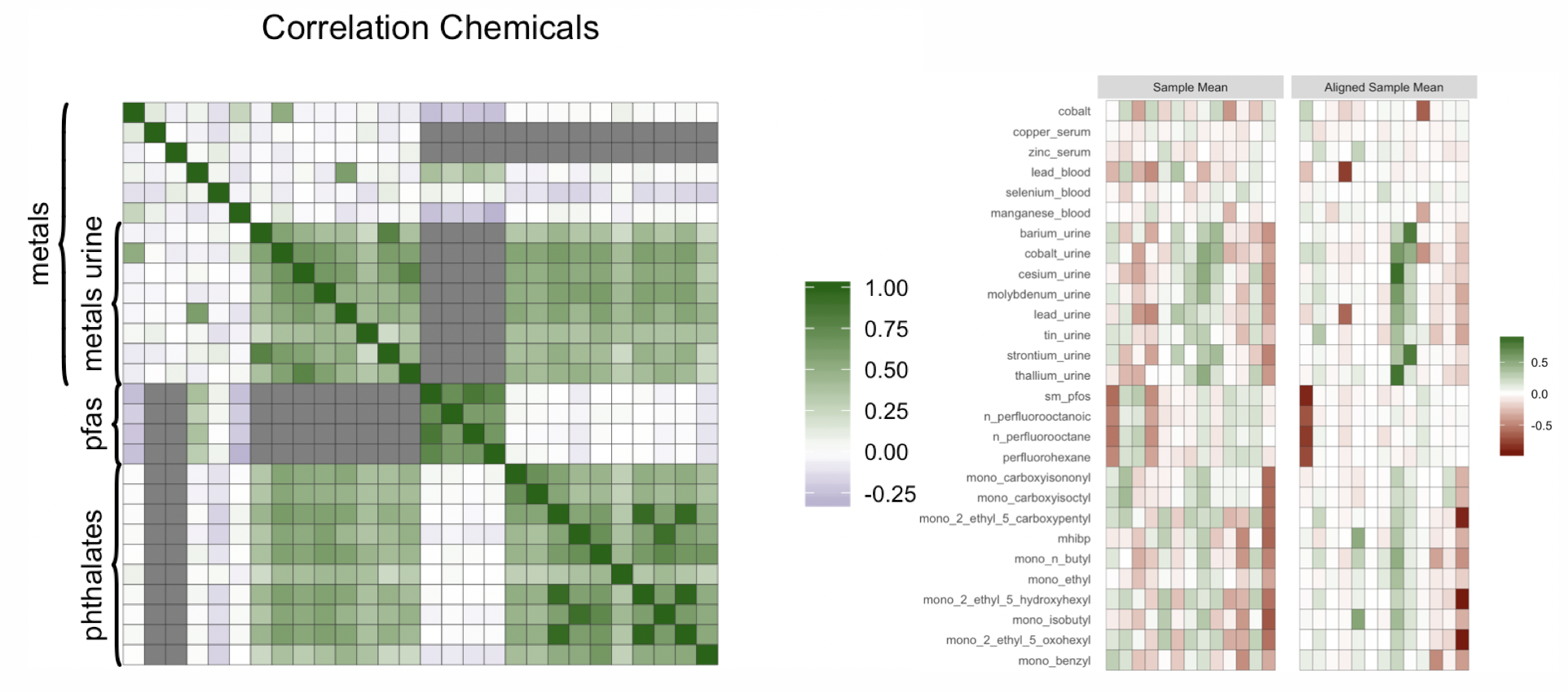}
\caption{On the left, correlation between the exposures, the colour grey indicates missing pairwise correlation. On the right, posterior mean of the matrix $\Lambda$ of factor loadings before and after applying the MatchAlign algorithm.}
\end{figure}

\begin{figure}[htbp]
\centering
\includegraphics[width=\linewidth, height= 6.5cm]{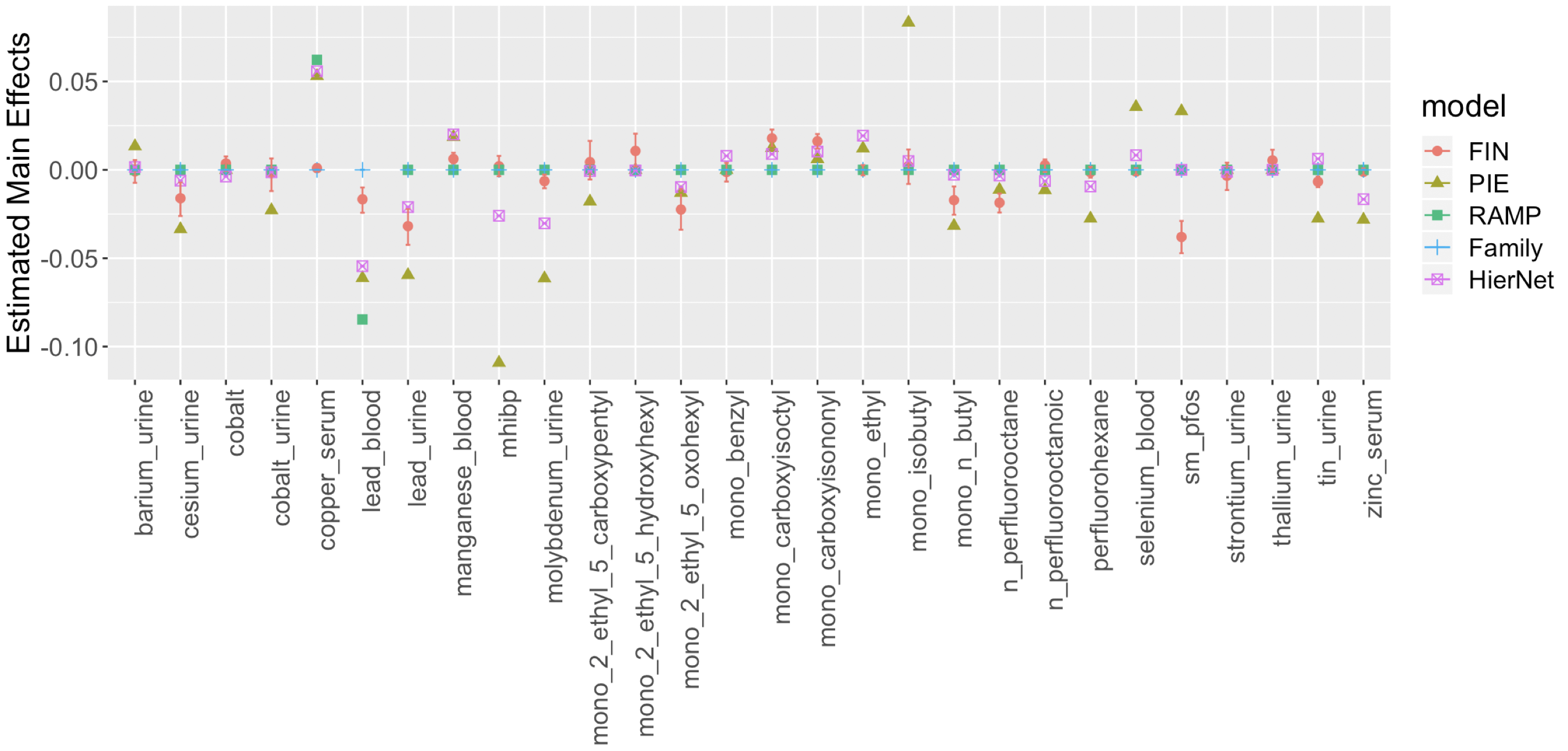}
\caption{Estimated main effects using FIN with $95\%$ credible intervals and estimated coefficients using RAMP, hierNet, Family and PIE.}
\end{figure}

\begin{figure}[htbp]
\begin{minipage}{0.47\textwidth}
\centering
\includegraphics[width=\linewidth, height= 6.5cm]{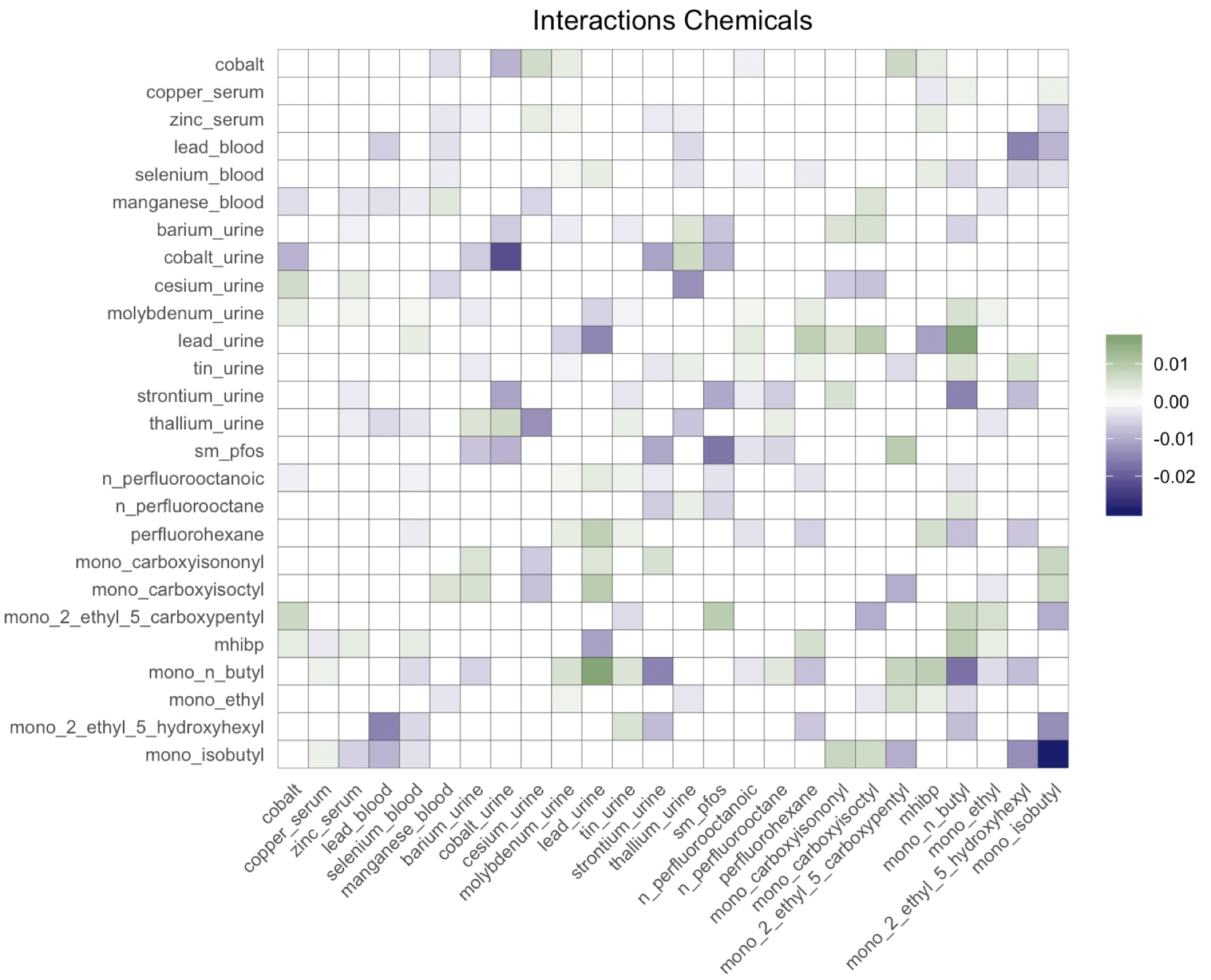}
\end{minipage} \hfill
\begin{minipage}{0.47\textwidth}
\centering
\includegraphics[width=\linewidth, height= 6.6cm]{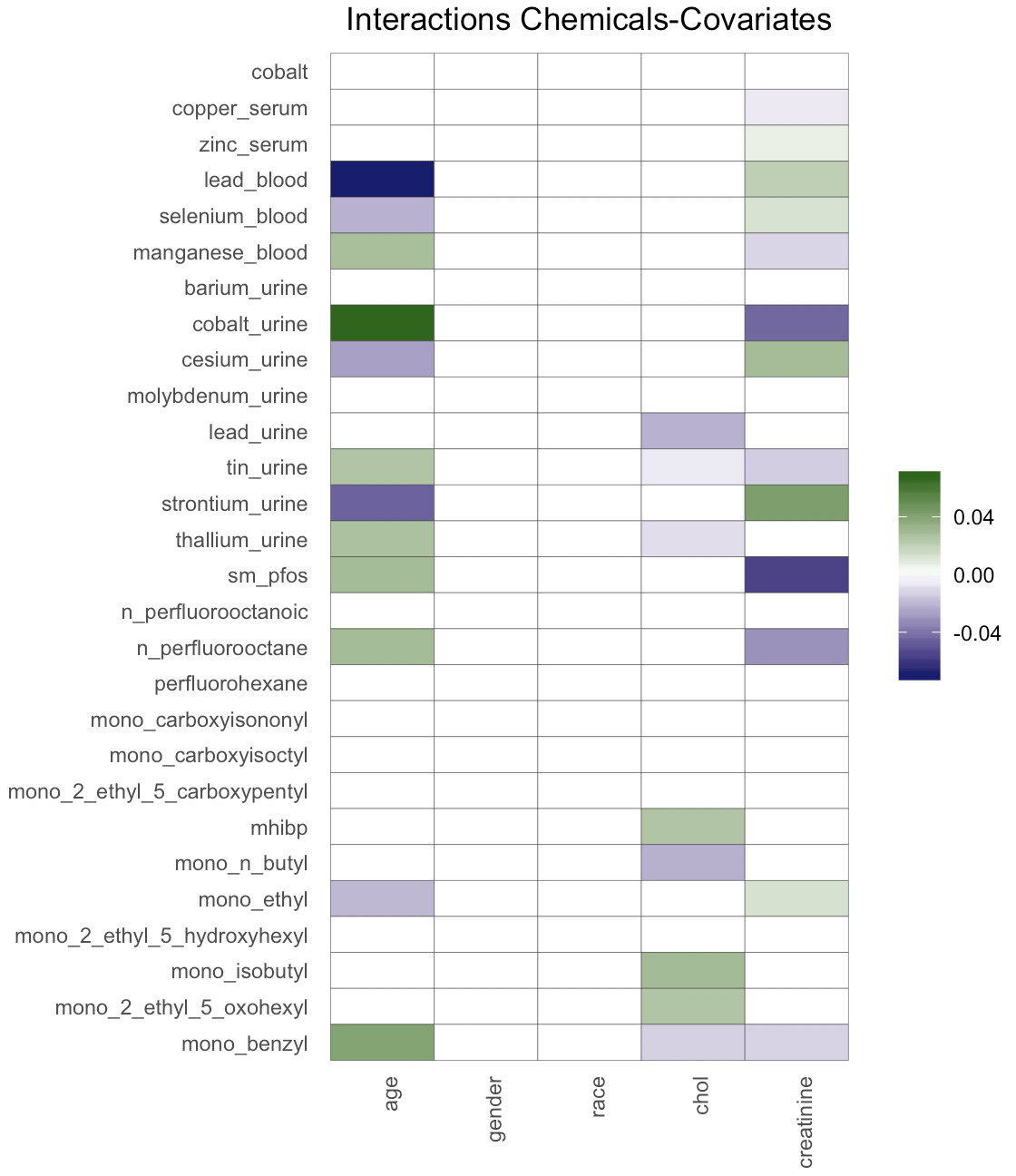}
\end{minipage}
\caption{On the left, posterior mean of the matrix of chemicals interactions. On the right, posterior mean of the matrix $A^T\Delta$ of pairwise interactions between exposures and covariates. The white boxes indicates that the $99\%$ credible interval contains zero.}
\end{figure}

\section{Discussion}

We proposed a novel method that exploits the correlation structure of the predictors and allows us to estimate interaction effects in high dimensional settings, assuming a latent factor model. Using simulated examples, we showed that our method has a similar performance to state-of-the-art methods for interaction estimation when dealing with independent covariates and outperforms the competitors when there is moderate to high correlation among the predictors. We provided a characterization of uncertainty with a Bayesian approach to inference. Our FIN approach is particularly motivated by epidemiology studies with correlated exposures, as illustrated using data from NHANES.

NHANES data are obtained using a complex sampling design, that includes oversampling of certain population subgroups, and contains sampling weights for each observation that are inversely proportional to the probability of begin sampled. We did not employ sampling weights in our analysis because our goal was to study the association between exposures and BMI rather than providing population estimates. One possibility to include the sampling weights in our method is to jointly model the outcome and the survey weights \citep{si2015bayesian}, without assuming that the population distribution of strata is known. 

Our MCMC algorithm can be efficiently employed for $n$ and $p$  in the order of thousands and hundreds respectively, which allows us to estimate around $5000$ interactions when $p = 100$. However, it is necessary to speed up the computations in order to apply our method to bigger $p$, which is common with genomics data. The computational bottleneck is the Metropolis Hastings step described in \textit{Section 2.2}. One possibility is to include the heredity constraint \citep{Chipman1996} while estimating the factors.

In order to allow departures from linearity and Gaussianity, it is of interest to model the regression on the health outcome as a non-linear function of latent factors. Non parametric latent models have desirable properties in term of convergence rates \citep{zhou2017adaptive} and large support for density estimation \citep{kundu2014latent}.  \cite{verma2018robust} developed a dimension reduction approach with latent variables for single cell RNA-seq data building on Gaussian process latent variable models (GP-LVM). Although attractive from a modeling perspective, a major challenge is efficient posterior computation. Another promising direction to decrease modeling assumptions is to rely on a copula factor model related to \cite{murray2013bayesian}.

\section*{Acknowledgments}

\if1\blind{
This research was supported by grant 1R01ES028804-01 of the National Institute of Environmental Health Sciences of the United States Institutes of Health. The authors would like to thank Evan Poworoznek, Antonio Canale, Michele Caprio, Amy Herring, Elena Colicino and Emanuele Aliverti for helpful comments.} \fi

\bibliographystyle{Chicago}

\bibliography{99BIB}

\section*{Appendix}



\begin{proof}[Proof of Proposition 1] \textit{(i)}
 Let us drop the $i$ index for notation simplicity and always assume that we are conditioning on all the parameters. The posterior distribution of $\eta$ is Normal with covariance $V = (\Lambda^T \Psi^{-1} \Lambda + I)^{-1}$ and mean $A X$ where $A = V  \Lambda^T \Psi^{-1}  = (\Lambda^T \Psi^{-1} \Lambda + I)^{-1} \Lambda^T \Psi^{-1}$. This follows from a simple application of Bayes Theorem. Now:
\begin{align*}
\E(y | X) & = \E(\E(y | \eta) | X) = \E(\eta^T \omega + \eta^T \Omega \eta | X  ) = \\
& = \omega^T \E(\eta |X ) + \E(\eta^T \Omega \eta |X) 
\end{align*}
Recall that the expectation of a quadratic form $\eta^T \Omega \eta$ of a random vector $\eta$ with mean $\mu$ and covariance matrix $\Sigma$ is equal to $tr(\Omega \Sigma) + \mu^T \Omega \mu^T$. 
\begin{align*}
\E(y | X)  & = \omega^T A X + tr (\Omega V_n) + (A X)^T \Omega (A X) = \\
& = tr (\Omega V)+(\omega^T A) X + X^T (A^T \Omega A) X
\end{align*}

\noindent \textit{(ii)} Recall that $\eta \sim N(0,I)$, $y = \eta^T \omega + \eta^T \Omega \eta_i +\epsilon_y$ and $X = \Lambda \eta+ \epsilon$, from simple algebra it follows that
\begin{align*}
Cov(y,X)  =  \omega^T Cov(\eta,\eta) \Lambda^T +   Cov(\eta^T \Omega \eta , \Lambda \eta )
\end{align*}
From the prior specification $Cov(\eta,\eta)=I$, hence let us focus on the term $Cov(\eta^T \Omega \eta , \Lambda \eta )$ and show that it is equal to $0_p$:
\begin{align*}
Cov(\eta^T \Omega \eta , \Lambda \eta ) & = Cov(\sum_{j = 1}^p \sum_{l=1}^p \omega_{j,l} \eta_j \eta_l , \Lambda \eta) = \\ 
& =\sum_{j = 1}^p \sum_{l=1}^p \omega_{j,l} Cov( \eta_j \eta_l ,\begin{pmatrix} \lambda_{1,1} \eta_1 + .... + \lambda_{1,k} \eta_k \\
... \\
\lambda_{p,1} \eta_1 + .... + \lambda_{p,k} \eta_k 
\end{pmatrix} ) =  \\
& = \sum_{j = 1}^p \sum_{l=1}^p \omega_{j,l} Cov( \eta_j \eta_l ,\begin{pmatrix} \lambda_{1,j} \eta_j + \lambda_{1,l} \eta_l \\
... \\
\lambda_{p,j} \eta_j + \lambda_{p,l} \eta_l
\end{pmatrix} ) = \\ 
& = \sum_{j = 1}^p \sum_{l=1}^p \omega_{j,l} \left[ Cov( \eta_j \eta_l ,\begin{pmatrix} \lambda_{1,j} \eta_j \\ ... \\ \lambda_{p,j} \eta_j  \end{pmatrix} ) +Cov( \eta_j \eta_l ,\begin{pmatrix} \lambda_{1,l} \eta_l \\ ... \\ \lambda_{p,l} \eta_l  \end{pmatrix} ) \right]
\end{align*}
Now $Cov(\eta_j \eta_l, \eta_j) = E(\eta_j^2 \eta_l) = 0$. In fact when $j \neq l $, we have that $E(\eta_j^2 \eta_l)  = E(\eta_j^2) E(\eta_l) = 0$ and when $j = l$, $E(\eta_j^3) = 0$ since $\eta_j \sim N(0,1)$.
\end{proof}

\end{document}